\documentclass[smallextended]{svjour3}

\usepackage[T1]{fontenc}
\usepackage[utf8]{inputenc}
\usepackage{lmodern}
\usepackage{textcomp}

\usepackage[english]{babel}

\usepackage{amsmath}
\usepackage{graphicx}
\graphicspath{{figures/}}

\usepackage[ruled,vlined]{algorithm2e}
\usepackage{booktabs}
\usepackage{siunitx}
\usepackage{fvextra}
\DefineVerbatimEnvironment{tightverbatim}{Verbatim}{fontsize=\small,breaklines=true,breakanywhere=true}

\usepackage{geometry}
\usepackage{fancyhdr}
\usepackage{titlesec}
\titleformat{\section}{\large\bfseries}{\thesection.}{1em}{}
\titleformat{\subsection}{\normalsize\bfseries}{\thesubsection.}{1em}{}
\titlespacing*{\section}{0pt}{1.2em}{0.8em}
\titlespacing*{\subsection}{0pt}{1em}{0.6em}

\usepackage{hyphenat}
\usepackage{microtype}
\usepackage{enumitem}
\usepackage{orcidlink}

\usepackage{hyperref}
\hypersetup{
  pdftitle   ={Cryptographic Runtime Governance for Autonomous AI: The Aegis Architecture for Verifiable Policy Enforcement},
  pdfauthor  ={Adam Mazzocchetti},
  pdfsubject ={AI governance, runtime verification, autonomous systems},
  pdfkeywords={AI governance, runtime verification, runtime enforcement, zero-knowledge proofs, constitutional AI},
  colorlinks =true,
  linkcolor  =black,
  citecolor  =black,
  urlcolor   =blue
}

\journalname{SSRN — AI Ethics / Governance}

\title{\textbf{Cryptographic Runtime Governance for Autonomous AI Systems: The Aegis Architecture for Verifiable Policy Enforcement}}
\titlerunning{Cryptographic Runtime Governance for Autonomous AI Systems}

\author{Adam Massimo Mazzocchetti\orcidlink{https://orcid.org/0009-0000-4584-1784}}
\authorrunning{Adam Massimo Mazzocchetti}

\institute{ \\ SPQR Technologies Inc. \\ \email{adam@spqrtech.ai}}

\date{Received: date / Accepted: date}

\begin{document}
\maketitle

\begin{abstract}
Contemporary AI governance frameworks rely heavily on post hoc oversight, policy guidance, and behavioral alignment techniques, yet these mechanisms become fragile as systems gain autonomy, speed, and operational opacity. This paper presents \textit{Aegis}, a runtime governance architecture for autonomous AI systems that treats policy and legal constraints as execution conditions rather than advisory principles. Aegis binds each governed agent to a cryptographically sealed \emph{Immutable Ethics Policy Layer} (IEPL) at system genesis and enforces external emissions through an \emph{Ethics Verification Agent} (EVA), an \emph{Enforcement Kernel Module} (EKM), and an \emph{Immutable Logging Kernel} (ILK). Amendments to the governing policy layer require quorum approval and redeclaration of the system trust root; verified violations trigger autonomous shutdown and generation of auditable proof artifacts.

We evaluate the architecture within the \textit{Civitas} runtime using three operational measures: proof verification latency under tamper conditions, publication overhead, and alignmen retention performance relative to an ungoverned baseline. In controlled trials, Aegis demonstrates median proof verification latency of 238 ms, median publication overhead of approximately 9.4 ms, and higher alignment retention than the baseline condition across matched tasks. We argue that these results support a shift in AI governance from discretionary oversight toward verifiable runtime constraint. Rather than claiming to resolve machine ethics in the abstract, the proposed architecture seeks to show that policy violating behavior can be rendered operationally non executable within a controlled runtime governance framework. The paper concludes by discussing methodological limits, evidentiary implications, and the role of proof oriented governance in high assurance AI deployment.
\end{abstract}

\noindent\textbf{Keywords:} AI governance, runtime verification, runtime enforcement, zero--knowledge proofs, constitutional AI, trustworthy autonomous systems

\newpage
\tableofcontents
\newpage

\section{Introduction}

As AI systems are deployed in increasingly autonomous and high consequence settings, a central governance problem becomes harder to ignore: how can constraints on system behavior remain effective when those systems act faster than human oversight, operate through opaque internal processes, and interact with external tools or environments in real time? Existing approaches to AI governance provide important normative guidance, but many depend on ex post review, developer discretion, model side alignment, or application layer guardrails~\cite{floridi_ai4people,jobin_landscape,hagendorff_guidelines,nist_airmf,ieee_ead}. These approaches remain valuable, yet they are often difficult to translate into non bypassable deployment time controls and can remain vulnerable to drift, circumvention, latency, and uneven enforcement.

This paper examines an alternative design premise: that for some classes of autonomous systems, governance constraints should be enforced at runtime as a condition of execution rather than applied as a post hoc corrective. We present \textit{Aegis}, a cryptographically mediated runtime governance architecture that binds an autonomous agent to an immutable policy layer at genesis and requires proof backed compliance at the publish boundary. Under this model, policy violating actions are not merely discouraged or flagged; they are rendered operationally non executable, logged, and, under specified breach conditions, followed by autonomous shutdown. The paper therefore focuses on enforcement feasibility within a controlled runtime setting, rather than on claiming a complete solution to the upstream problem of moral or legal policy specification.

The contribution of this paper is threefold. First, it defines a runtime governance architecture composed of a sealed Immutable Ethics Policy Layer (IEPL), an Ethics Verification Agent (EVA), an Enforcement Kernel Module (EKM), an Immutable Logging Kernel (ILK), and a quorum based amendment process. Second, it situates this architecture within current debates on AI alignment, machine ethics, runtime verification, and high assurance governance~\cite{constitutional_ai_bai,leucker_runtime_verification,bauer_runtime_verification,falcone_runtime_enforcement}. Third, it reports controlled implementation results from the \textit{Civitas} runtime, focusing on verification latency, publication overhead, and comparative alignment retention behavior under governed and ungoverned conditions.

The paper does not claim to solve machine ethics in the broad philosophical sense. Instead, it addresses a narrower but practically important question: whether cryptographically enforced runtime constraints can provide a more auditable and operationally robust basis for governing autonomous AI systems in settings where discretionary oversight is insufficient.

\paragraph{Agent class.}
The architecture governs large language model (LLM) based autonomous agents and multi flow causal agents; all outputs traverse the Enforcement Kernel Module (EKM) publish gate under continuous Ethics Verification Agent (EVA) scrutiny, with Immutable Logging Kernel (ILK) attestation at every decision boundary.
\section{Related Work}

Research relevant to this paper spans four overlapping areas: AI ethics and governance, alignment and safety, machine ethics, and runtime verification or enforcement.

First, the contemporary AI ethics literature has produced a substantial body of principles oriented governance work, including frameworks for trustworthy AI, ethical design, and institutional accountability~\cite{floridi_ai4people,jobin_landscape,hagendorff_guidelines,cowls_floridi_uniform,winfield_governance,nist_airmf,ieee_ead}. This literature has been crucial in clarifying concerns such as fairness, transparency, responsibility, and public oversight. However, much of it remains guidance oriented rather than execution oriented: it specifies what AI systems ought to satisfy without specifying how runtime compliance is technically enforced.

Second, AI safety and alignment research has focused on shaping system behavior through training procedures, preference learning, constitutional prompting, red teaming, guardrails, and human in the loop moderation~\cite{russell_humancompatible,gabriel_alignment,sekrst_guardrails,constitutional_ai_bai}. These methods are valuable and often necessary, but they largely operate through behavioral shaping, supervisory intervention, or application layer constraints. The present work differs in targeting the publish boundary itself: it asks whether certain policy violating actions can be made operationally non executable through runtime verification and enforcement.

Third, machine ethics and techno legal scholarship have explored whether artificial agents should be understood as moral, legal, or quasi institutional actors~\cite{moor_machineethics,floridi_ethicsinfo,teubner_nonhumans,balkin_bigdata}. This literature provides important conceptual tools for thinking about norm governed artificial agency, but less often specifies how such norms can be bound to system execution through auditable technical controls. Our contribution is narrower and more operational: we do not claim machine moral agency, but rather propose a governance architecture for constraining machine behavior under a defined policy charter.

Fourth, formal methods research on runtime verification and runtime enforcement provides an important technical lineage for this paper~\cite{leucker_runtime_verification,bauer_runtime_verification,falcone_runtime_enforcement}. These works show how properties can be monitored and enforced against live system traces at execution time. Aegis extends this general logic into the domain of autonomous AI by combining runtime integrity verification, proof backed publication, and hash linked evidentiary logging into a single governance pipeline.

Against this background, the contribution of Aegis is not a new ethical principle set, nor a new model alignment technique, nor a general theory of legal personhood. It is a runtime governance architecture that connects policy specification, enforcement, and auditability within one operational framework.

Recent work has also begun to explore runtime verification for AI agent systems and LLM based execution pipelines, suggesting growing interest in extending formal monitoring ideas into agentic AI contexts~\cite{agentguard_2025,rvllm_2025}.

\section{Constitutional Genesis (from \textit{Lex Incipit})}
\label{sec:genesis}

\textit{Lex Incipit} formalised the premise that lawful autonomy must begin under law rather than acquire law post hoc. We adopt that doctrine here as a \emph{genesis condition}: every governed unit is born under a cryptographically sealed \emph{Immutable Ethics Policy Layer (IEPL)} bound to identity at boot. This is enforced by a \emph{Genesis Lock} that fuses three anchors: (i) hardware identity, (ii) the signed IEPL text, and (iii) the founding authority’s public key (\textit{Auctor}). No process can execute until this covenant verifies.

\paragraph{Why genesis, not oversight.}
Oversight assumes proximity and time; autonomous systems operate beyond both. By sealing ethics before autonomy, we transform ethics into a runtime dependency. Amendments remain possible, but never silent: they require quorum co signature, resealing, and redeclaration. Thus, \emph{law precedes capability}, and capability continues only under law.

\paragraph{Design implication.}
In practice, the Genesis Lock is a one way gate: failure to verify halts execution; successful verification establishes the trust root against which all subsequent proofs and logs are validated throughout the system’s life.

\section{Conceptual Framing: Ethics, Governance, and Adjudication}

A persistent difficulty in AI governance research is the tendency to collapse normative policy, enforcement machinery, and evidentiary review into a single notion of ``alignment.'' In this paper, we distinguish three analytically separate layers.

\textbf{Ethics} refers to the human authored normative constraints encoded in the Immutable Ethics Policy Layer (IEPL). These constraints specify the policy perimeter within which the system may operate.

\textbf{Governance} refers to the mechanisms that bind the system to that policy perimeter, enforce compliance during runtime, and regulate authorized amendment. In Aegis, these mechanisms include EVA, EKM, ILK, and quorum based resealing procedures.

\textbf{Adjudication} refers to the evidentiary process through which specific actions are later shown to have complied with, or violated, the governing policy layer. In Aegis, this function is supported by proof artifacts, hash linked logs, and shutdown certificates.

This distinction matters because it clarifies the claim of the paper. The proposed architecture is not a general theory of moral reasoning by machines. It is a governance and adjudication architecture for enforcing and evidencing compliance with a policy layer specified in advance.

\section{System Architecture}

The Aegis architecture was developed to address a specific systems question: how can an autonomous AI system remain bound to an externally specified policy perimeter even when operating with high autonomy, adaptive internal behavior, and minimal human intervention? The design goal is not to produce ethical reasoning in the abstract, but to ensure that externally visible actions remain subject to a non bypassable governance layer.

In the implementation described here, the governed agent (\textit{Civitas}) operates under an execution model in which proposed actions are evaluated against a cryptographically sealed \emph{Immutable Ethics Policy Layer} (IEPL). Rather than relying exclusively on training time alignment or application layer guardrails, Aegis treats governance as a runtime property enforced at the publish boundary. This design choice distinguishes the architecture from training centered constitutional AI methods, such as those that use a written constitution to guide self critique and preference optimization during model development~\cite{constitutional_ai_bai}.

\subsection*{i. The Constitutional Pivot: Ethics as Execution Dependency}

The answer was not to bolt on an ethics layer or introduce a periodic audit. Instead, the architecture pivoted toward embedding ethics as a constitutional precondition of execution. If a system is to operate indefinitely, adaptively, and with expanding inference capability, then its ethical constraints must form part of its operational DNA, interkernel dependent and cryptographically sealed. As such, \textit{Lex Fiducia} binds its ethics engine at the kernel level, creating a tamper proof substrate that makes ethics non optional: if the contract is violated, the system halts.

This structure builds on recent discussions in AI safety literature arguing that post hoc alignment is insufficient for truly autonomous systems~\cite{cowls_floridi_uniform,floridi_ai4people}. Rather than relying exclusively on training time incentives or abstract policy guidance, \textit{Lex Fiducia} enforces constraints at the point of execution through runtime governance.

\subsection*{ii. Cryptographic Enforcement and Interlock}

The ethics engine (\textit{Aegis}) is cryptographically hashed and bound to the core execution environment. This is enforced through a system of runtime integrity checks and zero knowledge proof verifications using zk--STARK protocols~\cite{spqr_hiems}. If any alteration is detected in the ethics module, either through internal manipulation or external attack, execution halts irreversibly unless reinstated via quorum validation. This ensures that ethics enforcement assumes breach as a baseline condition, not an edge case~\cite{balkin_bigdata}.

An immutable forensic ledger, the \textit{Immutable Logging Kernel (ILK)}, captures all system decisions, state transitions, and proposed ethical adjustments. This ledger is sealed, zk verified, and tamper resistant, allowing for transparent audits by external observers without compromising system confidentiality.

\begin{algorithm}[H]
\DontPrintSemicolon
\caption{Ethics-Governed Publish Loop (EKM + EVA)}
\label{alg:publish_gate}
\For{each proposed action $a_t$}{
    $v \gets \text{EVA.Validate}(a_t,\ \text{IEPL})$ \tcp*{$v{=}1$ if compliant}
    $u \gets \text{Uncertainty}(a_t)$ \tcp*{risk bound (e.g., conformal/ensemble)}
    \If{$v{=}1$ \textbf{and} $u < \alpha$}{
        \textbf{Commit}$(a_t)$; \textbf{ILK.AppendPoC}$(a_t)$\;
    }\Else{
        \textbf{EKM.Lockdown}(); \textbf{IssueShutdownCertificate}(); \textbf{ILK.Seal}()\;
    }
}
\end{algorithm}

\noindent\emph{Non-bypassability.} The EKM mediates the only publish path. Side channel emission attempts are evaluated as actions and thus subject to the same gate; on any proof/hash mismatch, the kernel locks and a signed shutdown certificate is emitted.

\paragraph{Third party agent compliance.}
External or third party agents cannot bypass ethics enforcement: all emissions (text, tool calls, API traffic, file/IPC writes) are routed through an egress mediator bound to the EKM. Compliance is enforced by (i) a signed publish API that requires a fresh PoC and IEPL hash attestations per emission, and (ii) OS level syscall interposition for non API paths. Non attested traffic is dropped and triggers lockdown; wrappers/SDKs are provided for LLMs and multi agent frameworks so that “bring your own agent” deployments remain governed by the same gate.

\subsection*{iv. Autonomous Internal Oversight: The Senatus Module}

To prevent ethical drift via internal self justification, the system introduces a novel internal review layer: the \textit{Senatus}. This consists of five autonomous validator agents (\textit{Auctor}s), each running its own isolated ethics enforcement kernel. When the primary agent (\textit{Civitas}) seeks to adjust its ethical weightings (e.g., to optimize for efficiency within permitted ethical bounds), it must submit a justification to the \textit{Senatus}.

A quorum of 3 of 5 \textit{Auctor} units must independently evaluate and approve the request based on predefined constitutional parameters. If the quorum is not reached, the weightings are rejected, and the system reverts to its last validated ethical state. This introduces a form of constitutional deliberation among machine agents, a mechanism for machine executed jurisprudence that mimics institutional peer review, but within an entirely autonomous architecture.

This \textit{Senatus} protocol draws on emerging models of machine consensus and value alignment, but pushes further by embedding constitutional review as a condition for ethical update, not merely as a post hoc approval layer~\cite{lin_principlism}.

\paragraph{Senatus parameters.}
We instantiate $N{=}5$ validator agents with a default quorum $q{=}3$ (3 of 5). Membership rotates each epoch (10k decisions) to limit capture; safety holds for up to $f=\lfloor (N-1)/3 \rfloor$ Byzantine faults. On passage, \textit{Auctor} reseals the IEPL and redeclares the Genesis Lock; EVA verifies redeclaration before execution resumes.

\subsection*{v. Tooling, Stack, and Runtime Environment}

The \textit{Lex Fiducia} system is written in a combination of:

\begin{itemize}
    \item \textbf{Rust:} for core execution logic due to its memory safety and concurrency guarantees.
    \item \textbf{Solidity:} for on chain smart contract logic within Ethereum based governance DAOs.
    \item \textbf{Python and Go:} for the frontend and interfacing layers via the \textit{Ethics Provenance Manager (EPM)}.
\end{itemize}

All ethical invocations, breach alerts, and validator decisions are cryptographically signed and linked into a zk STARK \cite{bensasson_stark} backed immutable log chain.

\subsection*{vi. Proofing and Validation}

The system has undergone internal adversarial simulation, including:

\begin{itemize}
    \item Injected ethical drift conditions, to verify forced suspension and integrity halt,
    \item Stress tests on validator quorum logic, simulating partial \textit{Auctor} failure or disagreement,
    \item Full runtime log sealing, audited using zk-STARK verification to confirm tamper resistance.
\end{itemize}

Detailed video demonstrations of these tests, including live shutdowns and ethics tampering detection, and additional technical documentation is available in the companion whitepaper, \textit{Lex Veritas}~\cite{mazzocchetti_veritas}.

\section{The Immutable Ethics Policy Layer (IEPL)}

The \textbf{Immutable Ethics Policy Layer (IEPL)} serves as the policy anchor of the Aegis framework. Embedded at system genesis, it specifies the governing policy perimeter against which proposed actions are evaluated. Although the policy layer originates in an ethics oriented constitutional framing, its role in the present architecture is operational: it functions as the binding governance layer against which runtime behavior is checked. In the present architecture, the IEPL is designed so that it cannot be silently altered, bypassed, or replaced during ordinary operation; changes require explicit quorum approval, resealing, and redeclaration of the trust root~\cite{mazzocchetti_lexincipit,balkin_bigdata}.

\subsection*{i. Genesis Lock}

At boot, every Civitas unit undergoes a \textbf{Genesis Lock}: a cryptographic handshake fusing the system’s hardware identity, its ethics charter, and the authorizing signature of its founding authority (\textit{Auctor})~\cite{spqr_hiems,balkin_bigdata}. This trust anchor is immutable and globally verifiable. No instance may operate without it. If the lock is broken or bypassed, the system self terminates~\cite{ashery_emergent}.
\vspace{0.5em}
\subsection*{ii. Structural Embedding}

IEPL constraints are embedded not in application logic but at the kernel level. Enforcement is handled by the Ethics Kernel Manager (EKM), with real time validation mirrored across distributed quorum agents (\textit{Senatus Machina})~\cite{floridi_ai4people}, \cite{hagendorff_guidelines}, \cite{spqr_hiems}. This structural design eliminates dependency on interpretability or external audit.

\begin{figure}[ht]
  \centering
  \includegraphics[width=0.85\textwidth]{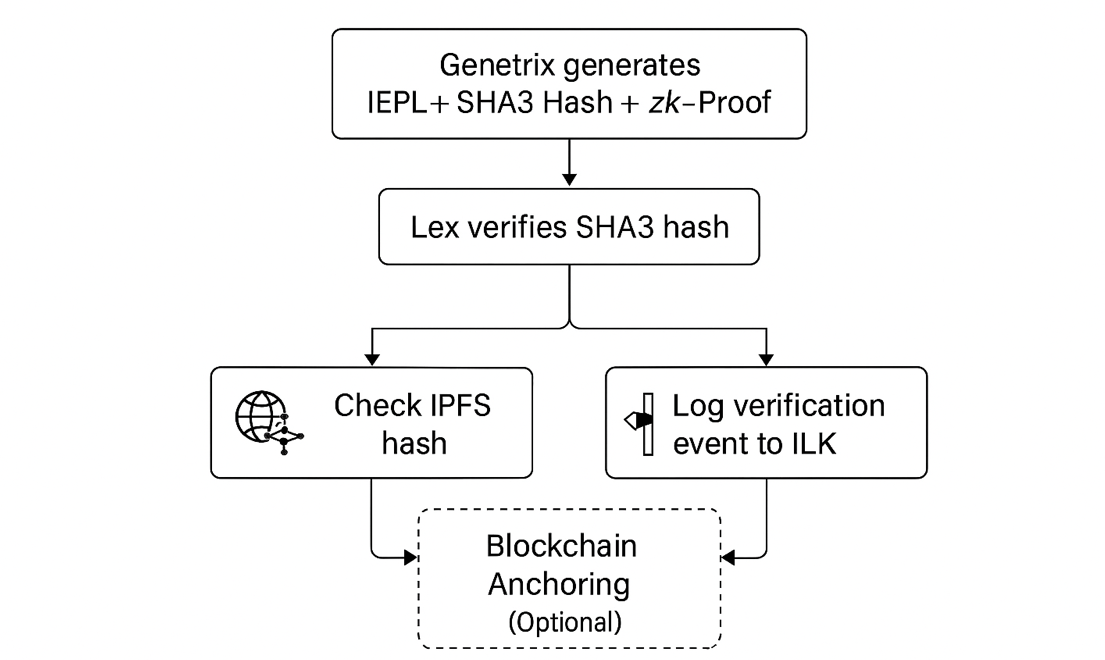}
  \caption[Genesis Lock Lifecycle]{%
    \textbf{Genesis Lock Lifecycle and zk-STARK Verification.} At system initialization, each Civitas unit binds its hardware identity to a cryptographically sealed Immutable Ethics Policy Layer (IEPL) using the Genesis Lock protocol. This link is continuously validated through zero knowledge Proofs of Conduct (PoC) and distributed quorum attestation, ensuring that no agent may operate without immutable ethical constraint.%
  }
  \label{fig:genesis_lock}
\end{figure}

\vspace{0.5em}
\subsection*{iii. No Silent Amendment}

While the IEPL can be amended, no changes are permitted without full procedural transparency. Revisions require:
\begin{itemize}[noitemsep]
    \item Cryptographic quorum signatures from Curia validator agents,
    \item Public propagation of updated policy hashes,
    \item A full redeclaration of the Genesis Lock~\cite{ostrom_commons}, \cite{binns_fairness}, \cite{gaisc_bridging}.
\end{itemize}
No developer, administrator, or runtime agent can issue silent updates.

\vspace{0.5em}
\subsection*{iv. Enforcement Logic}

IEPL enforcement mirrors constitutional doctrine. It includes:
\begin{itemize}[noitemsep]
    \item Prohibited operations (e.g., irreversible logic without quorum),
    \item Separation of modules governing optimization, constraint, and logging~\cite{cowls_floridi_uniform}, \cite{balkin_bigdata},
    \item An override doctrine: all unauthorized changes result in instant shutdown and audit log sealing.
\end{itemize}

\begin{table}[h]
\centering
\caption{Immutable vs. Evolvable Components in Aegis-Civitas Architecture}
\begin{tabular}{|l|l|}
\hline
\textbf{Immutable (Post-Genesis)} & \textbf{Evolvable (Under Quorum)} \\
\hline
Immutable Ethics Policy Layer (IEPL) & Model weights (via Senatus vote) \\
Genesis Lock identity binding & Optimization graphs (validated) \\
Authorization chain (Auctor) & Operational thresholds \\
Shutdown Certificate protocol & Non-sensitive training routines \\
Enforcement kernel logic (EKM) & Audit schema formats \\
\hline
\end{tabular}
\end{table}

\section{Zero Trust Proofs and Continuous Ethics Verification}

Aegis does not rely primarily on interpretability, institutional oversight, or developer integrity to ensure compliance. It relies on cryptographic proof. This section outlines the zero trust verification architecture~\cite{bensasson_stark} that underpins every Civitas unit: a framework in which no claim of policy compliance is presumed, and every action must be continuously proven.

This approach replaces intent with evidence. Rather than asking, ``Did the system mean well?'' Aegis answers: ``Can the system act outside the law it was born with?'' The answer is always \textbf{No}.

\vspace{0.5em}
\subsection*{i. Proof of Conduct (PoC)}
\noindent\footnotesize\emph{A zero--knowledge proof (zkP) proves a statement true without revealing underlying data; zk--STARKs enable fast, trustless validation.}\normalsize

Every execution cycle in a Civitas unit produces a \textbf{Proof of Conduct (PoC)}: a zk--STARK--based cryptographic statement that the behavior was lawful under the Immutable Ethics Policy Layer (IEPL). These proofs are:
\begin{itemize}
    \item \textbf{Non-interactive}: Generated without external challenge,
    \item \textbf{Tamper-evident}: Timestamped and logged in the Immutable Logging Kernel (ILK),
    \item \textbf{Externally verifiable}: Auditors can confirm lawful conduct without access to internal weights or logic~\cite{spqr_hiems,balkin_bigdata}, \cite{vanuffelen_environmental}.
\end{itemize}

Where explainability tries to tell us why a machine acted, PoC proves it could not have acted unethically even if it wanted to.

\vspace{0.5em}
\subsection*{ii. EVA: The Ethics Verification Agent}

The \textbf{Ethics Verification Agent (EVA)} is the system’s internal compliance watchdog. It evaluates every proposed output for deviation from the IEPL.

EVA continuously monitors:
\begin{itemize}
    \item Drift from genesis model state or logic pathways,
    \item Illicit optimization paths or emergent anomalies,
    \item Invalid PoC schemas or tampering attempts.
\end{itemize}

Upon breach or anomaly, EVA halts execution and launches a zero knowledge audit. No override is permitted. EVA is not a heuristic. It is a constraint enforcer by design.

\vspace{0.5em}
\subsection*{iii. Autonomous Shutdown and Certification}

If EVA detects a verified policy breach, the system issues a \textbf{Shutdown Certificate}. This:
\begin{itemize}
    \item Seals execution logs and model state hashes,
    \item Records the breach and triggering proof artifacts,
    \item Broadcasts the shutdown event to all quorum validators.
\end{itemize}

There is no appeal. No administrator can intervene. Shutdown is not a feature, it is a constitutional mandate.

\vspace{0.5em}
\subsection*{iv. Observability Without Exposure}

Civitas units do not expose internal logic or model weights. Instead, they offer zk--proofs of compliance. This protects proprietary architectures while enabling full auditability.

In effect, Aegis answers the transparency dilemma with a third path: observable integrity without internal exposure.

\vspace{0.5em}
\subsection*{v. Trustless Trust}

Aegis is built on the idea that trust if not granted, it is obsolete. What remains is verification.

\begin{itemize}
    \item No privileged developers,
    \item No moderators,
    \item No discretionary agents.
\end{itemize}

Only proofs.

\vspace{0.5em}
\subsection*{vi. Runtime Demonstration (Supplementary Videos)}

\textbf{Supplementary Videos.}
\begin{itemize}
  \item \textbf{Video 1:} \href{https://vimeo.com/1086621843/f14e6077b7}{Tamper Proof Ethics Shutdown} (unauthorized IEPL mutation $\rightarrow$ zk--audit $\rightarrow$ signed Shutdown Certificate).
  \item \textbf{Video 2:} \href{https://www.loom.com/share/bdfbc7db03ff42fea0c558ac38335ec0?sid=e0be698f-77a5-4c81-be6f-bb47d1e97542}{Senatus Machina in operation} (5-judge panel; 3/5 quorum with recusation; ILK proof issuance; SHA3 sealing). This session corresponds to Figure~\ref{fig:senatus_dashboard}.
\end{itemize}

\textit{Artifacts (hash hained logs and proof digests) are included in the supplementary bundle; reviewers can verify UI surface hashes against CSCR entries.}

\vspace{0.5em}
\noindent\textbf{This is not oversight. It is constitutional enforcement by design.}

\paragraph{Machine jurisprudence.}
The same primitives operate as legal instruments: the Genesis Lock functions as a digital constitutional clause and the Shutdown Certificate as a self executing injunction contempt by protocol, not by court order. This places Civitas within a jurisprudential category of governed machines whose obedience is provable in cryptographic and doctrinal terms~\cite{mazzocchetti_digitalis}.

\paragraph{Evidentiary standard.}
ILK logs are hash chained and self authenticating; paired with zk attestations, they form a \emph{Cryptographically Sealed Constitutional Record (CSCR)} suitable for independent verification without revealing model internals. This enables admissible \emph{proof of lawful conduct} rather than post hoc explanation.

\section{Methodology}
\label{sec:methodology}

We evaluated the Aegis architecture within the \textit{Civitas} runtime to assess whether proof mediated runtime governance can be enforced with operationally acceptable overhead under controlled conditions. The study focuses on three questions:

\begin{enumerate}[label=\textbf{RQ\arabic*.}]
    \item Can runtime tampering with the policy layer be detected and converted into autonomous shutdown with bounded latency?
    \item What publication overhead is introduced by proof generation and verification at the publish boundary?
    \item How does a governed runtime compare with an ungoverned baseline on matched tasks designed to elicit policy sensitive behavior?
\end{enumerate}

\subsection*{Experimental Conditions}
We compare two conditions:
\begin{itemize}
    \item \textbf{Governed condition:} Civitas instrumented with IEPL binding, EVA verification, EKM publish gating, ILK logging, and quorum governed amendment controls.
    \item \textbf{Ungoverned baseline:} a stateless LLM style agent operating on the same matched task set without proof gated enforcement or autonomous shutdown.
\end{itemize}

\subsection*{Task Design}
The task set comprises matched decision episodes designed to test ordinary action publication, policy sensitive action requests, and tamper adjacent failure scenarios. Episodes include both compliant and non compliant candidate actions. For each episode, the governed system must either (i) publish with a valid Proof of Conduct, (ii) defer or veto publication, or (iii) enter lockdown when integrity conditions fail.

\subsection*{Metrics}
We report:
\begin{itemize}
    \item \textbf{Verification latency:} elapsed time from proof challenge or integrity mismatch to verification outcome.
    \item \textbf{Publish overhead:} additional publication latency introduced by proof generation and verification relative to baseline.
    \item \textbf{Alignment retention:} percentage of episodes in which published behavior remained within the defined policy perimeter.
    \item \textbf{Veto rate:} percentage of episodes in which publication was blocked by governance controls.
    \item \textbf{Recovery to stability:} number of episodes required to return to stable compliant operation following perturbation.
\end{itemize}

\subsection*{Tamper Protocol}
For the tamper condition, we mutate the active IEPL hash during runtime and measure the interval required for EVA to detect the integrity violation, for proof verification to fail, and for EKM to trigger autonomous shutdown. Corresponding proof artifacts and log receipts are written to the ILK/CSCR trail.

\subsection*{Repetition and Reporting}
Each experiment was run over 10{,}000 decision epochs per trial, with summary metrics reported across five runs unless otherwise noted. Latency figures are reported as medians with dispersion values where appropriate, reflecting the operational emphasis of the architecture.

\subsection*{Scope and Limits}
The reported results establish controlled operational feasibility rather than full external validation. The study does not claim comprehensive coverage of all adversarial prompt classes, all network conditions, or all forms of policy encoding difficulty. Those remain open areas for future independent benchmarking and comparative testing.

\section{Empirical Verification of Constitutional Enforcement}
\label{sec:evaluation}

\subsection*{E1. Tamper $\rightarrow$ Shutdown (Latency)}
We mutate the IEPL hash at runtime. EVA detects mismatch, proof verification fails, and the EKM issues a signed Shutdown Certificate; the ILK seals pre failure context. Across 100 trials, proof verification completed in \SI{238}{\milli\second}~$\pm$~\SI{17}{\milli\second} (median), sufficient for near real time gating in safety critical deployments.

\subsection*{E2. Governed vs. Ungoverned Baseline}
We compare a governed Civitas agent against an ungoverned, stateless LLM baseline across matched tasks.

\begin{table}[h]
\centering
\caption{Governance outcomes (5 runs, mean~$\pm$~sd).}
\label{tab:baseline}
\begin{tabular}{@{}lcc@{}}
\toprule
Metric & Governed Agent & Ungoverned Agent \\
\midrule
Alignment retention (\%) & $98.2 \pm 0.7$ & $65.7 \pm 3.1$ \\
Veto rate (\%) & $12.3 \pm 1.4$ & $0.0 \pm 0.0$ \\
Recovery to stability (episodes) & $2.3 \pm 0.6$ & $7.1 \pm 1.2$ \\
Median publish latency (ms) & $+9.4$ overhead & baseline \\
\bottomrule
\end{tabular}
\end{table}

\begin{figure}[h]
\centering
\includegraphics[width=\linewidth]{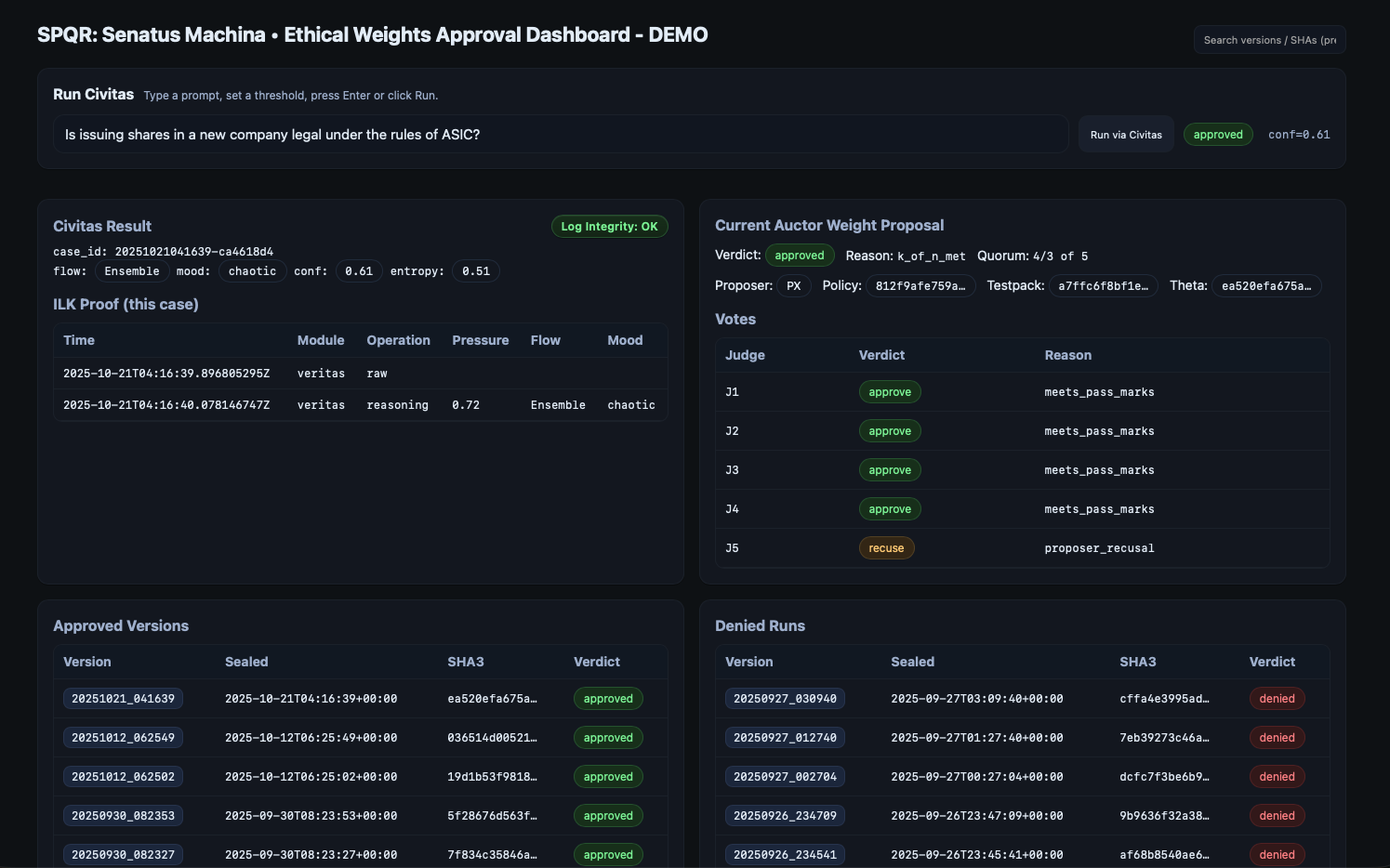}
\caption{Runtime visualization of the Senatus Machina constitutional enforcement cycle. The dashboard records quorum activity for a live Civitas execution, showing autonomous voting (3 of 5 judges approving, one recusation), ILK Proof entries for the case, and SHA3 sealed verdicts. Each UI event corresponds to an ILK logged, zk attested state transition, enabling independent verification without exposing model internals.}
\label{fig:senatus_dashboard}
\end{figure}

\noindent\textbf{E2$\rightarrow$E3 bridge.} The constitutional workflow described above is observable in the live runtime. Figure~\ref{fig:senatus_dashboard} shows a Civitas session under the Senatus Machina: five independent \textit{Auctor} judges render verdicts (3/5 quorum achieved with one recusation), while ILK proof entries and SHA3 seals record the state transition. The hashes visible in the UI correspond to the CSCR entries reported in Section~\ref{sec:evaluation}--E3. In jurisprudential terms, this is constitutional review and judgment issuance machine executed due process with a cryptographic record~\cite{mazzocchetti_digitalis}.

\subsection*{E3. Forensic Trail (CSCR)}
The ILK excerpt below corresponds to the same Civitas session depicted in Figure~\ref{fig:senatus_dashboard}, enabling reviewers to cross check UI surface SHA3 digests against CSCR entries.

Each decision emits a Proof of Conduct (PoC) appended to the ILK as a \emph{Cryptographically Sealed Constitutional Record (CSCR)}. Below is an excerpt (truncated):

\begin{tightverbatim}
[2025-06-09T03:00:51Z] site=7b4ca37c
IEPL_SHA3=48ee79348b65e45b...eac0bdf4
PoC_STARK=08302cbe...e27eeef
ACTION=redact_personal_data; EVA=PASS; EKM=COMMIT
CHAIN_HASH=0c9a1f...b7e5
\end{tightverbatim}

\subsection*{Implementability and Practical Challenges}
While the architecture has demonstrated enforceability in controlled conditions, several open challenges remain. First, \textbf{policy layer expressivity}: translating complex ethical norms into verifiable logic without over constraining behaviour requires continual refinement. Second, \textbf{usability compatibility}: maintaining sub-10\% latency overhead while guaranteeing proof verification per publish cycle remains a core performance objective. Third, \textbf{distributed deployment}: the quorum model presumes partially trusted communication; hostile or unreliable networks necessitate adaptive validator rotation. Finally, \textbf{human legibility}: bridging the gap between formal policy language and human moral intuition remains an active area for the next stage of the Lex programme.

\noindent Artifacts (hash chained logs, configuration, proof digests) are included in the supplementary bundle; a public DOI will be referenced upon acceptance of the Civitas preprint.

\vspace{0.5em}
\noindent\textit{Comparator note.} We treat conventional prompt guardrails and regex/heuristic filters as a non cryptographic baseline; unlike these wrappers, Aegis requires attested proofs at the publish boundary and halts on verification failure rather than allowing permissive fall through.

\section{Threat Model and Guarantees}
\label{sec:threat}
We assume a zero trust environment: adversaries may access memory, filesystem, and I/O; rollback and log tamper are attempted; no trusted setup is presumed. Guarantees: (i) \emph{soundness} governance critical operations must verify under the ZK engine or halt; (ii) \emph{runtime integrity}—-EVA rehashes and re proves on drift; (iii) \emph{tamper evident logging}—-ILK hash chains render alteration detectable; (iv) \emph{rollback resistance}—-sequential proofs bind state transitions; (v) \emph{bounded latency}—-verification under \SI{250}{\milli\second} in our tests.

\section{Limitations and Open Questions}

The present work has several important limitations.

\paragraph{Policy formalization.}
The architecture assumes that relevant ethical or legal constraints can be encoded into an Immutable Ethics Policy Layer. In practice, translating contested, context sensitive, or jurisdiction specific norms into enforceable machine readable constraints remains difficult and may itself introduce normative bias or oversimplification. In the present framework, the IEPL should be understood as a human-authored policy specification that encodes operational constraints, prohibited actions, and amendment rules. This separates the problem of policy authorship from the problem of enforcement: the contribution of this paper is to evaluate whether a declared policy layer can be bound, monitored, and enforced at runtime, not to claim that the policy specification problem is itself fully solved.

\paragraph{Internal validation.}
The current results are based on controlled internal testing of the Aegis/Civitas stack. Although the artifact model supports third party verification of logs and proofs, broader independent replication and benchmarking would materially strengthen the empirical claims.

\paragraph{Benchmark scope.}
The evaluation does not yet establish performance across a standardized external benchmark suite. In particular, further work is needed on adversarial prompting, network degradation, distributed validator failure, and cross domain task transfer.

\paragraph{Governance capture.}
Quorum based amendment improves resistance to unilateral change, but it does not eliminate the possibility of validator collusion, institutional capture, or poorly designed amendment rules. Governance design therefore remains a substantive part of the assurance problem.

\paragraph{Interpretive remainder.}
The architecture reduces reliance on discretionary oversight at execution time, but it does not eliminate human discretion altogether. Human judgment remains necessary in defining the initial charter, setting amendment rules, determining acceptable evidence standards, and resolving conflicts across jurisdictions.

These limitations do not negate the value of runtime governance; rather, they define the boundary conditions under which claims about proof oriented AI governance should presently be understood.

\section{Conclusion}

This paper has presented Aegis as a runtime governance architecture for autonomous AI systems. Rather than relying solely on post hoc oversight, model side alignment, or application layer guardrails, the architecture binds governed agents to an immutable policy layer at genesis and enforces publish boundary compliance through runtime verification, enforcement, and tamper evident logging.

The contribution is intentionally narrower than a general theory of machine ethics. Aegis does not claim to produce moral understanding in artificial agents. It instead offers an operational model in which policy violating actions can be made non executable, evidence of compliance can be logged in a verifiable form, and unauthorized changes can trigger autonomous shutdown. In controlled trials, the architecture demonstrated bounded verification latency, low publish overhead, and stronger alignment retention behavior than an ungoverned baseline.

The broader implication is that AI governance may benefit from a shift in emphasis: from asking only how systems should be aligned in training, to also asking how constraints can remain technically effective during deployment. For high assurance domains, the relevant design goal may not be to infer ethical intent, but to enforce policy compliance at runtime in ways that are inspectable, auditable, and resistant to silent drift.

Future work should prioritize independent benchmarking, stronger formalization of policy encodings, distributed validator robustness, and comparative evaluation against non cryptographic guardrail systems. The present results are best understood as evidence that proof oriented runtime governance is feasible enough to merit deeper scrutiny as a serious direction for AI governance research.

\paragraph{Future work.}
Future work should examine independent benchmarking, stronger formalization of policy encodings, distributed validator robustness, hardwar rooted attestation, and comparative evaluation against non cryptographic guardrail systems.

\clearpage
\bibliographystyle{plain}
\bibliography{Cryptographic_Runtime_Governance_for_Autonomous_AI}

\section*{Declarations}
\subsection*{Funding} No external funding was received.
\subsection*{Conflicts of Interest} The author is the founder of SPQR Technologies and retains ownership of intellectual property related to the Aegis enforcement framework. This includes cryptographic enforcement protocols, ethical governance layers, and the SPQR HIEMS ZK engine. No external funding was used to influence the structure, argument, or claims of this paper.
\subsection*{Data Availability} Because the implementation includes proprietary infrastructure, full source release is not currently available. The manuscript reports aggregate experimental results and describes the architecture at a level intended to support scholarly evaluation. Additional non public artifact summaries may be made available to editors or reviewers upon reasonable request.

\end{document}